\documentclass[aps,prb,showpacs,groupedaddress,superscriptaddress,twocolumn]{revtex4}

\usepackage{graphicx,amsmath}
\begin{document}

\title{Surface Step Defect in 3D Topological Insulators: Electric Manipulation
of Spin and Quantum Spin Hall Effect}

\author{Yan-Feng Zhou}
\affiliation{International Center for Quantum Materials, School of Physics, Peking University, Beijing 100871, China}
\affiliation{Collaborative Innovation Center of Quantum Matter, Beijing 100871, China}

\author{Ai-Min Guo}
\affiliation{Department of Physics, Harbin Institute of Technology, Harbin 150001, China}

\author{Qing-Feng Sun}
\email[]{sunqf@pku.edu.cn}
\affiliation{International Center for Quantum Materials, School of Physics, Peking University, Beijing 100871, China}
\affiliation{Collaborative Innovation Center of Quantum Matter, Beijing 100871, China}

\date{\today}

\begin{abstract}
We study the influence of step defect on surface states in three-dimensional topological insulators subject to a perpendicular magnetic field. By calculating the energy spectrum of the surface states, we find that Landau levels (LLs) can form on flat regions of the surface and are distant from the step defect, and several subbands emerge at side surface of the step defect. The subband which connects to the two zeroth LLs is spin-polarized and chiral. In particular, when the electron transports along the side surface, the electron spin direction can be manipulated arbitrarily by gate voltage. And no reflection occurs even if the electron spin direction is changed. This provides a fascinating avenue to control the electron spin easily and coherently. In addition, regarding the subbands with high LL index, there exist spin-momentum locking helical states and the quantum spin Hall effect can appear.
\end{abstract}


\maketitle

\section{\label{sec1}Introduction}

In recent years, many effects have been achieved on topological insulators (TIs)\cite{HMZ,Qxl}, which are an intriguing insulating phase of electronic materials and can be classified by different topological invariants\cite{TDJ,KCL}. The electronic states at surface, namely surface states, of three-dimensional (3D) TIs are conducting and are protected by time-reversal symmetry\cite{Fl1,MJE,Fl2,HD,Chen,Xia}, which arises from the bulk topology through bulk-boundary correspondence\cite{Fl2}. For the 3D TIs, the excitations of the surface states are characterized by 2D Weyl fermions and present an odd number of gapless Dirac cones in the Brillouin zone. In contrast to conventional two-dimensional (2D) conductors, the conducting surface states are robust against nonmagnetic disorders due to the prohibited backscattering and could induce the half-quantum Hall effect under magnetic field.

Several experimental\cite{ZT,AZ,WJ} and theoretical studies\cite{BRR,ZD,An} have demonstrated the Friedel oscillations of local density of states, which is attributed to scattering from surface step in the 3D TIs and provides evidence for suppression of backscattering. The Friedel oscillations present power-law decay behavior with exponent ranging from -3/2 to -1/2, by either changing the Fermi energy or extending step defect along different directions, due to strong warping effect in the 3D TIs\cite{An,WJ}. In these theoretical studies, the step defect is regarded as potential barrier of either one-dimensional $\delta$ function\cite{ZD,An} or step function\cite{WJ}, where the incident electron is perpendicular to the step defect. If the electron transports parallel to the step defect, this method cannot capture all the physics, since all the surfaces construct a closed one\cite{Lee,VO}. Recently, the matching conditions are derived for the surface states of two perpendicularly connected surfaces of the 3D TIs\cite{BL}. By using these matching conditions, one can study the electronic structure of the surface states of the step defect in the 3D TIs.

Since the surface states possess a novel helical spin texture, they can be used in the field of spintronics\cite{HMZ,LQ,YT,GI}. The experiments have realized the injection and detection of spin-polarized current in the 3D TIs by electric methods\cite{Li,TJ,AY,LL,TJF,DA}. Another important theme of spintronics is to coherently manipulate the spin degree of freedom. The spin polarization of the surface states can be tuned by polarized illumination\cite{MT,JC,YD}. In principle, the spin polarization direction can be modulated by external magnetic field and by optical methods, but it is difficult to produce integrated circuits with low power consumption\cite{ZL,ADD}. So, it is important to control the electron spin by using the electric methods, just as the Datta-Das transistor\cite{Datta,KHC,CP}, which is a spin field effect transistor and the spin direction is tuned by gate voltage with the aid of Rashba spin-orbit coupling.

In this paper, we study the influence of the step defect on the surface states in the 3D TIs subject to a perpendicular magnetic field ($z$-axis), as illustrated in Fig.1(a). Under the magnetic field, Landau levels (LLs) can form in both upper and lower surfaces, and are distant from the step defect (side surface). Although the side surface is parallel to the magnetic field, its surface states can develop into subbands due to the size quantization effect. And the expectation of the electron spin is calculated for these subbands. Our results reveal that the subband which connects to the two zeroth LLs is spin polarized and chiral. When the electron transmits within the side surface, the spin polarized direction can be tuned by gate voltage, and no reflection occurs by changing the electron spin direction. These unique features provide a new avenue to control the electron spin easily and coherently. Besides, the subbands with nonzero indices are parabolic and present helical property, which generates the quantum spin Hall effect and is similar to the helical edge states in the 2D TIs.

The rest of the paper is organized as follows. In Sec.~\ref{sec2}, the theoretical model and the methods for studying the influence of the step defect are presented.
In Sec.~\ref{sec3}, we propose an electric strategy to manipulate the electron spin.
In Sec.~\ref{sec4}, we show a quantum spin Hall effect in a convex platform
on the TI's surface at zero gate voltage.
Finally, a brief summary is given in Sec.~\ref{sec5}.

\section{\label{sec2}Model and Methods}

\begin{figure}
\includegraphics[scale=0.5]{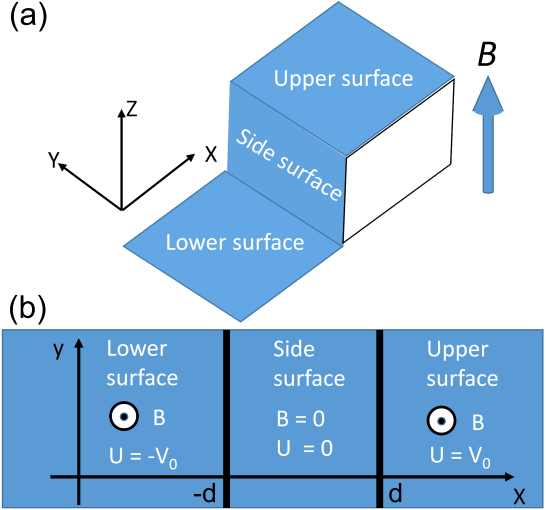}
\caption{ (Color online) (a) Schematic of upper and lower surfaces connected by side surface (step defect) in the 3D TI under perpendicular magnetic field. (b) Mapping of the above surfaces in the regions of $x>d$, $-d<x<d$, and $x<-d$ of 2D model. The potential energies of the upper and lower surfaces can be controlled by the gate voltage and no magnetic field threads into the side surface.}
\end{figure}

The 3D TI is composed of the upper surface, the lower one, and the side one (step defect), as shown in Fig.1(a). Then, the effective Hamiltonian of the surface states can be expressed as\cite{LHZ,Liu1}
\begin{equation}
H({\bf k}) = \hbar \nu_{F}(\hat{\sigma}\times {\bf k})\cdot \hat{n} +U,
\end{equation}
where $\nu_{F}$ is the Fermi velocity, $\hat{\sigma} \equiv (\sigma_x,\sigma_y,\sigma_z)$ with $\sigma_{x,y,z}$ the Pauli matrices, ${\bf k}=(k_x,k_y,k_z)$ is the momentum, $\hat{n}$ is the normal vector of a specific surface, and $U$ is the potential energy which can be modulated in the experiments by the gate voltage. Here, we assume that $\nu_{F}$ is the same for all the three surfaces mentioned above. For the side surface, we make a local rotation by fixing the $y$-axis so that the original $x$-axis is changed into the $z$-axis, and the effective Hamiltonian of the side surface can be obtained from a unitary transformation. By combining the matching conditions between the side surface and the $+z$ surface, the effective 2D model can be obtained and then the three surfaces can be described by the Hamiltonian $H=\hbar\nu_{F}(-k_x \sigma_y+ k_y\sigma_x) +U(x)$.\cite{TM,TM2}

In this effective model, the upper, lower, and side surfaces locate in the regions of $x>d$, $x<-d$, and $-d< x < d$, respectively, and are infinite along the $y$-axis [Fig.1(b)]. Under the Landau gauge, the vector potential can be written as $A = (0, A_{y})$, with $A_{y}$ being $B(x-d)$, 0, and $B(x+d)$ for the upper, side, and lower surfaces, respectively, where the parameter $B$ is the strength of the magnetic field. We consider that the gate electrodes only locate on the top of the upper and lower surfaces, and the potential energies are $U(x) =V_{0}$, $0$, and $-V_0$, respectively, in the upper, side, and lower surfaces. As the model is invariant by translating along the $y$-axis, the momentum $k_{y}$ is a good quantum number.

To calculate the band structure of the 2D model, the effective Hamiltonian can be discretized along the $x$-axis by performing $\frac{d\Psi(x)}{dx}\rightarrow \frac{\Psi_{\mathrm{i}+1}-\Psi_{\mathrm{i}-1}}{2a}$:\cite{book1}
\begin{equation}\label{eq:12}
\begin{split}
&H = \sum \limits_{\mathrm{i}}[c_{\mathrm{i}}^{\dag}T_{0}c_{\mathrm{i}}
   + (c_{\mathrm{i}}^{\dag}T_{x}c_{\mathrm{i}+1} + H.c.)],\\
   &T_{0}= U(x_{\mathrm{i}})I+(W/a)\sigma_{z}+\nu_{F}(\hbar k_{y}+eA_{y})\sigma_{x},\\
   &T_{x}=-(W/2a)\sigma_{z}+(i\hbar \nu_{F}/2a)\sigma_{y},\\
\end{split}
\end{equation}
where $c_{\mathrm{i}}$ and $c_{\mathrm{i}}^{\dag}$ are, respectively, the annihilation and creation operators at site $\mathrm{i}$. $I$ is the $2\times2$ unit matrix and $W/a$ is the Wilson mass term which is introduced to avoid the fermion double problem, with $a$ the lattice constant and $W$ being $0.1\hbar \nu_{F} $.\cite{Kogut}
Here, the Zeeman effect is not considered, because some experiments have shown that the g-factor is quite small
and the Zeeman effect may not be important.\cite{Zeeman1,Zeeman2}

\begin{figure}
\includegraphics[scale=0.3]{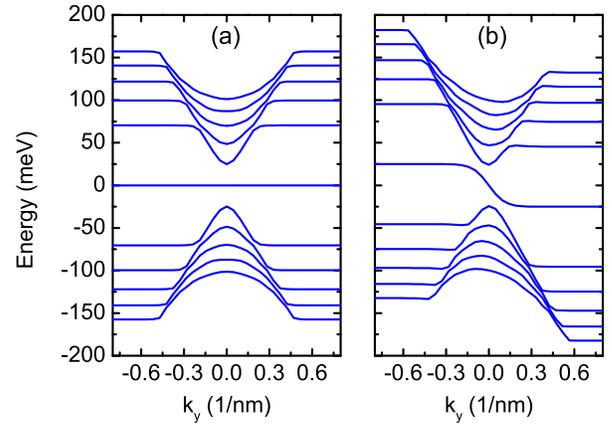}
\caption{(Color online). Dispersion relation of the 2D model with (a) $V_{0} = 0 \ \mathrm{meV}$ and (b) $V_{0} = 25 \ \mathrm{meV}$. The other parameters are $2d = 30 \ \mathrm{nm}$ and $B = 15\ \mathrm{T}$.} \label{fig:2}
\end{figure}

\section{\label{sec3}Manipulation of electron spin}

Firstly, we calculate the band structure of the 2D effective model, where the Fermi velocity is set to $\nu_{F} = 5\times10^{5}\ \rm m/s$,\cite{ZT,YR} the lattice constant is $a = 0.375\ \mathrm{nm}$, and the truncation length along the $x$-axis is taken as $180 \ \mathrm{nm}$, which is sufficiently long that the wavefunctions of the LLs are localized. For $V_{0} = 0\ \mathrm{meV}$, the 2D model has $C_2$ symmetry with respect to the $z$-axis. As a result, the band structure is symmetric, i.e., $E(k_{y})= E(-k_{y})$, as indicated in Fig.2(a). It is clear that except for the zero energy band which is flat, each band consists of two flat parts and a parabola-like part. In the flat parts, the states are the LLs with the energy $\varepsilon_{N}= \mathrm{sgn}(N) \hbar\omega_{c} \sqrt{|N|}$, and correspond to the upper surface for $k_{y}<0$ and to the lower surface for $k_{y}>0$, where the integer number $N$ is the LL index and $\omega_{c}=\nu_{F}\sqrt{2eB/\hbar}$ is the cyclotron frequency. While in the parabola-like parts, the states are confined to the side surface and form subbands owing to the size quantization effect. When the gate voltage is implemented, the band structure becomes asymmetric and the energies of the LLs for the upper (lower) surface are shifted by $V_{0}$ ($-V_{0}$) [see Fig.2(b)]. And the band connecting to the two zeroth LLs becomes chiral, similar to the chiral edge state in the topological p-n junction\cite{Wj}.

\begin{figure}
\includegraphics[scale=0.3]{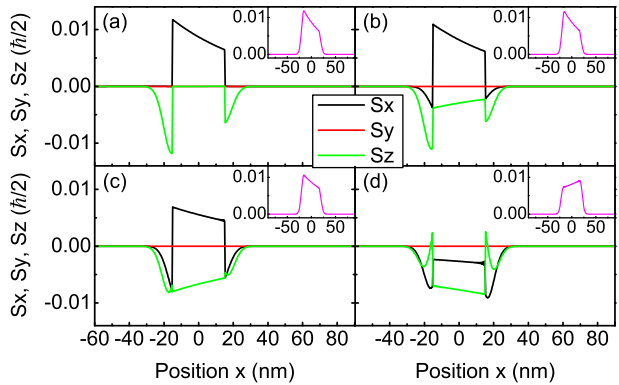}
\caption{ (Color online) Spatial distribution $\mathbf{s}_ {x,y,z}$ of the electron spin for the chiral state, which connects to the two zeroth LLs of the upper and lower surfaces, with the potential energy (a) $V_{0}= 0\ \mathrm{meV}$, (b) $V_{0} = 10\ \mathrm{meV}$, (c) $V_{0} = 25\ \mathrm{meV}$, and (d) $V_{0} = 50\ \mathrm{meV}$. The insets display probability density $\rho(x)$. Here, we choose $k_{y}=0.01\ \mathrm{nm^{-1}}$, $2d = 30\ \mathrm{nm}$, and $B = 15 \  \mathrm{T}$.} \label{fig:3}
\end{figure}

We then study the property of the chiral state which connects to the two zeroth LLs of the upper and lower surfaces. Fig.3 plots spatial distribution $\mathbf{s}_{x,y,z}$ of the electron spin for the chiral state with different values of $V_{0}$, and the insets show the corresponding probability density $\rho(x)$ versus position $x$. Here, $\mathbf{s}_{x,y,z} = \Psi^{\dag}_{N,k_y}(x) \mathbf{\hat{s}}_{x,y,z}\Psi_{N,k_y}(x)$, $\rho(x)=|\Psi _{N,k_y}(x)|^2$, and the expectation of the electron spin can be obtained by the integral $\mathbf{\bar{s}}_{x,y,z}= \int \mathbf{s}_{x,y,z}dx$, where $\Psi_{N,k_y}(x)$ is the eigenfunction of Eq.(2) and $N$ is the band index. As mentioned above, the 3D model [Fig.1(a)] can be mapped into the 2D effective model [Fig.1(b)] by performing local rotation on the side surface, where the $x$-$z$ plane is rotated by $90^{\circ}$ with the $y$-axis fixed. Because of this local rotation, the spin operator is expressed as $\mathbf{\hat{s}}= \frac{\hbar}{2}(\sigma_{x}, \sigma_{y}, \sigma_{z})$ when $x > d$ and $x < -d$, and $\mathbf{\hat{s}}= \frac{\hbar}{2} (-\sigma_{z}, \sigma_{y}, \sigma_{x})$ when $-d < x < d$.

When $k_y$ is close to $0$, the chiral state mainly locates in the side surface (see the insets of Fig.3). It can be seen from Fig.3(a) that in the region of the side surface, the chiral state is spin-polarized along the $+x$ direction, although the magnetic field is parallel to the side surface and has no direct effect on it. While beyond the side surface, the electron spin is polarized along the $-z$ direction. It is worth noting that $\mathbf{s}_{y}$ is always zero for all the bands because the Hamiltonian is real.
When the gate voltage is implemented, a negative $\mathbf{s}_{z}$ occurs in the side surface [see Figs.3(b)-3(d)].
In the region of the side surface, $\mathbf{s}_{x}$ decreases with $V_{0}$ and even becomes negative for large $V_{0}$, and the absolute value of $\mathbf{s}_{z}$ is enhanced.
This implies that the electron spin direction can be changed from $\mathbf{s}_x$-up to $\mathbf{s}_x$-down.
The underlying physics for the
change of the spin direction is that the spin-orbit coupling can modify the two
components of the eigenstates in the surface when the gate voltage is changed.
Besides, the chiral state still resides in the side surface. Therefore, this provides a strategy to manipulate the electron spin direction coherently by tuning the gate voltage.

\begin{figure}
\includegraphics[scale=0.33]{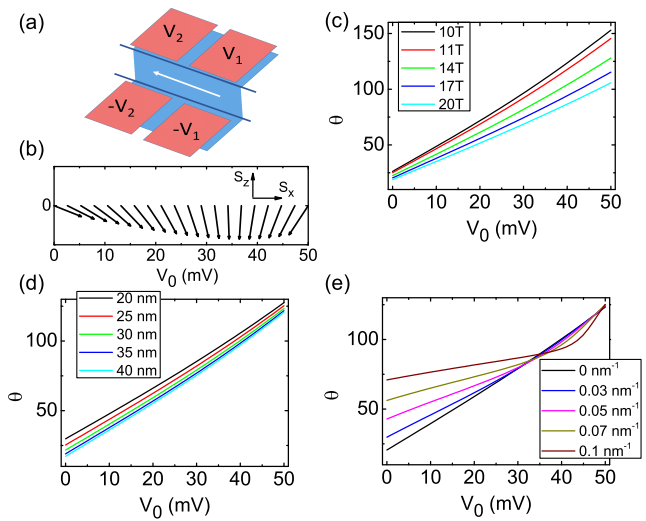}
\caption{ (Color online) (a) A setup for manipulation of the electron spin under nonuniform gate voltage. (b) Spin texture in the $\mathbf{s}_x$-$\mathbf{s}_z$ plane as a function of the gate voltage $V_0$. Tilting angle $\theta$ versus $V_{0}$ for (c) different magnetic fields $B$, for (d) various heights $2d$ of the side surface, and for (e) several momentums $k_y$. The unmentioned parameters are $B = 15\ \mathrm{T}$, $2d = 30\ \mathrm{nm}$, and $k_y=0.01\ \mathrm{nm^{-1}}$. } \label{fig:4}
\end{figure}

To further demonstrate the details of this strategy, we study the spin polarization of the chiral state under the gate voltage for different system parameters. Fig.4(b) shows spin texture of the chiral state versus the gate voltage $V_0$.
The arrow in the spin texture denotes the spin polarization vector of an eigenstate at a specific gate voltage. The spin polarization vector is defined as $\mathbf{P}=(\mathbf{\bar{s}}_x,\mathbf{\bar{s}}_y,\mathbf{\bar{s}}_z)$, where the $\mathbf{\bar{s}}_{x,y,z}$ are the expectations of the spin operator as mentioned above. It clearly appears that the electron spin direction can be successively tuned from $\mathbf{s}_{x}$-up to $\mathbf{s}_{x}$-down by increasing $V_{0}$. To quantitatively analyze the electron spin, a tilting angle is defined as $\theta = \arccos[\mathbf {\bar{s}}_{x} /(\mathbf{\bar{s}}_{x}^2 +\mathbf{\bar{s}}_ {z}^2) ^{1/2}]$, which denotes the relative orientation of the electron spin direction with respect to the $x$-axis. Figs.4(c) and 4(d) present the tilting angle $\theta$ for different magnetic fields $B$ and for different heights $2d$ of the side surface, respectively, as a function of $V_0$. It can be seen that $\theta$ increases almost linearly with $V_0$ for different magnetic fields $B$ and heights $2d$. For a specific height $2d$, the weaker the magnetic field is, the faster $\theta$ changes with $V_0$ [Fig.4(c)]. While for a certain magnetic field, the slope of the $\theta-V_{0}$ curve is only slightly modified by changing $2d$ [Fig.4(d)]. The tilting angle $\theta$ is slightly increased by decreasing $2d$, which is irrespective of $V_0$. The range of the tilting angle $\theta$ becomes small
by decreasing $2d$.
When the height $2d$ is declined to zero, the system becomes a topological p-n junction and the range of $\theta$ will be less than $65^{\circ}$ by changing $V_0$ from $0$ to $50$ meV.
This implies that the existence of the side surface is important to manipulate the spin direction,
because the spin direction in the side surface is very different from that in the upper and lower surfaces.
Do all the chiral states in Fig.2(b) possess these characters? By inspecting Fig.4(e), one can see that for a wide range of $k_y$ in which the band is chiral, the tilting angle $\theta$ rises monotonously as $V_0$ increases. There exists an obvious deviation from the linear behavior when the states approach the LLs.

Thus, it is feasible and effective to manipulate the electron spin of the chiral states by the gate voltage. Based on the above results, we can envision a setup [see Fig.4(a)], in which several independent gates are placed on the top of both upper and lower surfaces, and the electron transports along the side surface. By tuning the gate voltage, the potential energies can be set arbitrarily in space and the electron spin direction can be modulated in any way. Furthermore, there is not any reflection during the transport process because of the unidirectionality of the chiral state in the side surface. In a word, this is a high controllable, no reflection, low energy consumption, and electric strategy to manipulate the electron spin direction.

\begin{figure}
\includegraphics[scale=0.3]{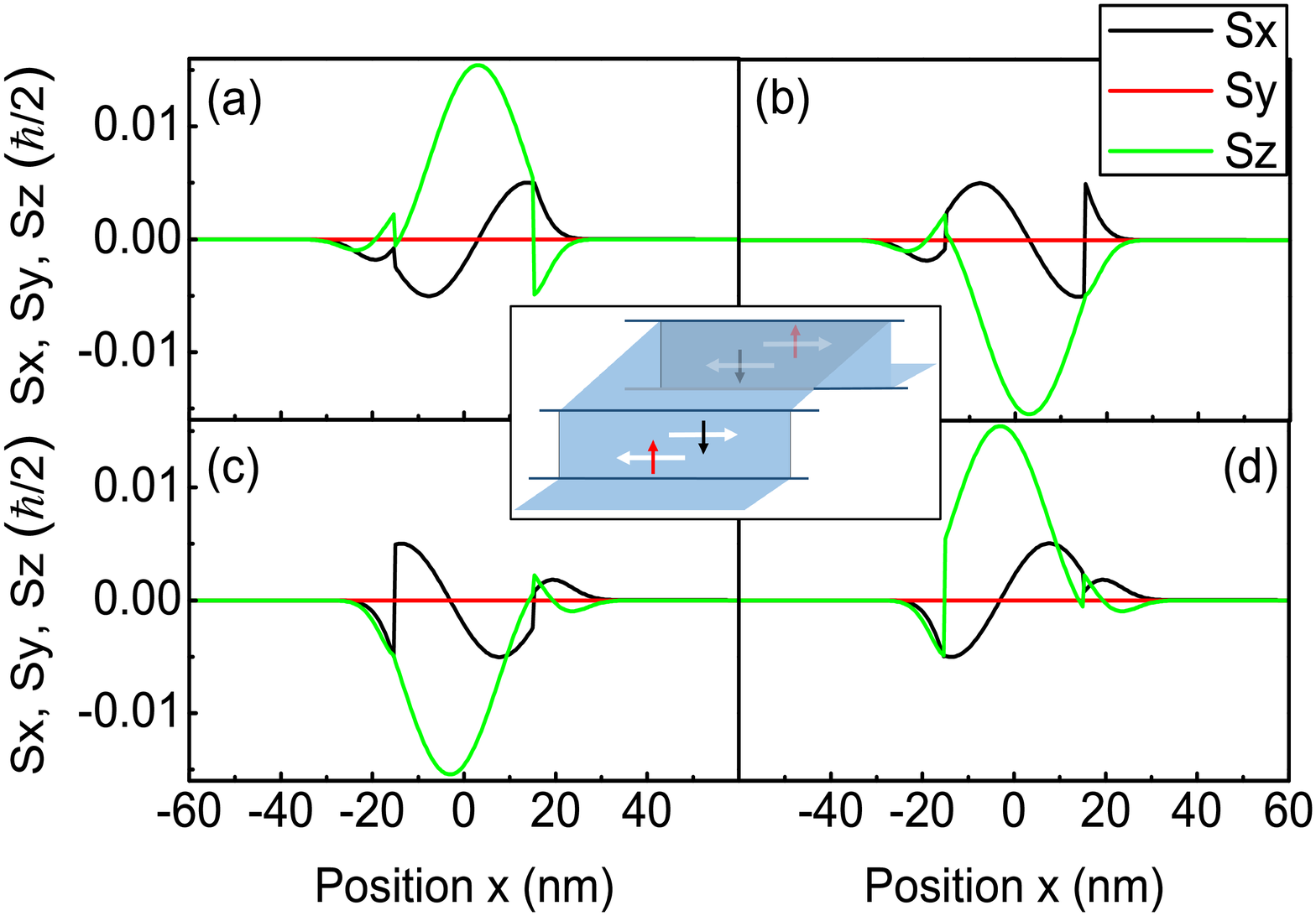}
\caption{ (Color online) (a) and (c) Spatial distribution $\mathbf{s}_{x,y,z}$ for the electronic state in the front side surface, which connects the two LLs of $N=1$, with (a) $k_y=0.1\ \mathrm{nm^{-1}}$ and (c) $k_y=-0.1\ \mathrm{nm^{-1}}$. (b) and (d) show the corresponding $\mathbf{s}_{x,y,z}$ for the back side surface. The parameters are $B = 15\ \mathrm{T}$, $2d = 30\ \mathrm{nm}$, and $V_0=0\ \mathrm{meV}$. In the calculation of (b) and (d), we make local rotation of the $x$ and $z$ coordinates around the $y$-axis by $-90^{\circ}$, and the region of $-d<x<d$ is still the side surface. The inset shows schematic view of a convex platform in the surface of the 3D TI, and the front and back side surfaces form naturally. } \label{fig:5}
\end{figure}

\section{\label{sec4}Quantum spin Hall effect}

Finally, we consider the 3D TI with a convex platform on its top surface, as depicted in the inset of Fig.5, and study the bands which connect to the nonzero LLs of the upper and lower surfaces. In this 3D TI, the front and back side surfaces appear naturally at the boundary of the convex platform and are of identical height. The distance between the two side surfaces is sufficiently long so that the localized states in the two side surfaces cannot mix together under the perpendicular magnetic field. The magnetic field is set to $B = 15\ \mathrm{T}$, the height of the side surfaces is $2d = 30\ \mathrm{nm}$, and no gate voltage is employed.

The electronic states connecting to the nonzero LLs are non-chiral near the front side surface, and there are a couple of states with the same energy but propagating oppositely in the side surface [see Fig.2(a)]. Figs.5(a) and 5(c) display the spatial distribution $\mathbf{s}_{x,y,z}$ for the states connecting to the $N=1$ LLs with $k_y = \pm 0.1\ \mathrm{nm^ {-1}}$. It is evident that the state with $k_y=0.1\ \mathrm{nm^ {-1}}$ is propagating forward and is almost polarized in the $+z$ direction [Fig.5(a)], while the state with $k_y=-0.1\ \mathrm{nm^ {-1}}$ is propagating backward and is polarized in the $-z$ direction [Fig.5(c)]. The spatial distribution $\mathbf{s}_y$ is exactly zero everywhere and $\mathbf{\bar{s}}_x$ is very small. Therefore, these states in the front side surface are spin-momentum locking helical states, which are similar to the helical edge states in the 2D TI\cite{KCL,Jiang1}.

While for the back side surface, the forward state is polarized in the $-z$ direction [Fig.5(b)] and the backward state is polarized in the $+z$ direction [Fig.5(d)]. Therefore, the spin-momentum locking persists for these electronic states. And the quantum spin Hall effect can be observed in the convex platform of the 3D TI under the magnetic field, and the longitudinal conductance will be quantized when the Fermi energy lies within the gap between the two LLs of the upper and lower surfaces. Besides, the bands of higher LLs are also spin-momentum locked (data not shown). Thus, by tuning the Fermi energy, the quantum spin Hall effect can be achieved with $N$ spin-up states propagating forward and $N$ spin-down states propagating backward. And some quantum plateaus can be detected in the longitudinal resistance and the spin Hall resistance\cite{Jiang1,Sun1}.

\section{\label{sec5}Conclusions}

In summary, we have investigated the effect of step defect on surface states in three-dimensional topological insulators under a perpendicular magnetic field. By calculating the energy spectrum and the expectation of the electron spin of the surface states in the zeroth band, a chiral subband exists in the side surface and the electron spin direction can be manipulated in any way by gate voltages on the flat regions. Besides, no reflection occurs when the electron spin direction of the chiral state is changed. This provides a low power and electric scheme to control the electron spin coherently. In addition, for each subband with high indices, there is a pair of spin-momentum locking helical states in the side surface, and these helical states can induce quantum spin Hall effect.

\section*{Acknowledgments}

We gratefully acknowledge the financial support from NBRP of China (2012CB921303 and 2015CB921102), NSF-China under Grants No. 11274364, 11574007, and 11504066, and FRFCU under Grant No. AUGA5710013615. Y.-F. Zhou gratefully acknowledges T. Morimoto for many helpful discussions.

\end{document}